\documentclass[%
 reprint,
 amsmath,amssymb,
 aps,
]{revtex4-2}

\usepackage{graphicx}
\usepackage{dcolumn}
\usepackage{bm}
\usepackage[dvipsnames]{xcolor}

\begin{document}

\preprint{APS/123-QED}

\title{Steady-State Ultracold Plasma}

\author{B. B. Zelener}
\affiliation{Joint Institute for High Temperatures, Russian Academy of Sciences, Moscow 125412, Russia}
\email[]{bobozel@mail.ru}
\author{E. V. Vilshanskaya}
\affiliation{Joint Institute for High Temperatures, Russian Academy of Sciences, Moscow 125412, Russia}
\author{N. V. Morozov}
\affiliation{Joint Institute for High Temperatures, Russian Academy of Sciences, Moscow 125412, Russia}
\author{S. A. Saakyan}
\affiliation{Joint Institute for High Temperatures, Russian Academy of Sciences, Moscow 125412, Russia}
\author{A. A. Bobrov}
\affiliation{Joint Institute for High Temperatures, Russian Academy of Sciences, Moscow 125412, Russia}
\author{V. A. Sautenkov}
\affiliation{Joint Institute for High Temperatures, Russian Academy of Sciences, Moscow 125412, Russia}
\author{B. V. Zelener}
\affiliation{Joint Institute for High Temperatures, Russian Academy of Sciences, Moscow 125412, Russia}

\date{\today}

\begin{abstract}
  A strongly coupled ultracold plasma can be used as an excellent test platform for studying many-body interactions associated with various plasma phenomena. In this paper we discuss an approach that makes possible creation of the steady-state ultracold plasma having various densities and temperatures by means of continuous two-step optical excitation of calcium atoms in the MOT. The parameters of the plasma are studied using laser-induced fluorescence of calcium ions. The experimental results are well described by a simple theoretical model involving equilibration of the continuous source of charged particles by the hydrodynamical ion outflux and three-body recombination. The strongly coupled plasma with the peak ion density of $6\cdot10^5$~cm$^{-3}$ and the minimum electron temperature near 2~K has been prepared. Our steady-state approach in combination with the strong magnetic confinement of the plasma will make it possible to reach extremely strong coupling in such system.
\end{abstract}

\maketitle


Since its creation for the first time, ultracold plasma~\cite{killianCreationUltracoldNeutral1999} has proven to be an excellent tool for experimental study of various phenomena associated with low-temperature plasma physics. In particular, this kind of experiments made it possible to directly test many theoretical approaches to plasma such as treatment by means of the Vlasov equations, the three-body recombination theory, electron-ion equilibration rate estimations, magnetic field influence and others; for more details see comprehensive reviews~\cite{bergesonExploringCrossoverHighenergydensity2019,killianUltracoldNeutralPlasmas2007,lyonUltracoldNeutralPlasmas2017}.

An important feature of an ultracold plasma is its ability to achieve strong coupling. The coupling strength of a plasma is determined by the ratio of the mean interaction energy of the plasma particles to their mean kinetic energy. Analytical estimations and numerical calculations demonstrate that the ionic component of strongly coupled ultracold plasma may exhibit a short-range order or even a long-range order~\cite{bonitzThermodynamicsCorrelationFunctions2004,pohlCoulombCrystallizationExpanding2004a}. Numerical simulations show that the experiments involving an ultracold plasma can simulate more convenient hot plasma with the same coupling strength~\cite{bergesonExploringCrossoverHighenergydensity2019,bobrovConductivityDiffusionCoefficients2019}. In addition, strongly coupled ultracold plasma can be used as an ultracold ion source for scanning ion microscopy~\cite{jacobTransmissionMicroscopyNanometer2016}.

Most research of ultracold plasma (UCP) is focused on a high-density UCP with relatively short lifetime, which usually does not exceed 100~$\mu$s~\cite{killianCreationUltracoldNeutral1999,bergesonExploringCrossoverHighenergydensity2019}. Recent experiments involving laser cooling of ions~\cite{langinLaserCoolingIons2019} and magnetic confinement of UCP~\cite{gormanMagneticConfinementUltracold2021} made it possible to increase the lifetime and coupling strength of an ultracold plasma, but the lifetime remained limited. However, there are some applications which require studying properties of strongly coupled plasma whose lifetime is unlimited, i.e. a plasma in a steady-state mode. Such plasma is formed, for example, in case of continuous real-time injection of a lithium aerosol into the edge plasma in tokamaks~\cite{huNewSteadyStateQuiescent2015} or due to plasma transport in a rotation-dominated magnetosphere of Jupiter containing an internal plasma source from its satellite Io~\cite{pontiusjr.SteadyStatePlasma1986}.

In our study we overcame the problem of a limited lifetime of an UCP. Here we report on an experiment which results in obtaining a continuous source of an UCP in operating magneto-optical trap~(MOT) by ionizing of the trapped atoms using a continuous wave~(cw) laser beam. In this way, we make continuous conversion of calcium atoms from the magneto-optical trap to the ultracold plasma. As a result, the steady-state ultracold plasma~(SUCP) is formed in the MOT region in about a hundred of milliseconds after the ionizing laser is switched on. This novel ultracold plasma mode combines strong coupling and infinite time of existence. 

\begin{figure}
  \includegraphics[width=0.48\textwidth]{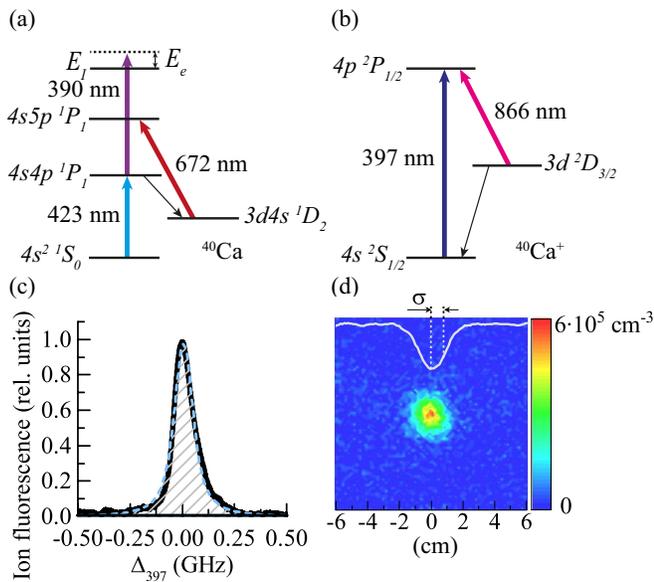}
  \caption{\label{fig:1} (a)~Energy levels of the $^{40}$Ca atom; $E_I$ is ionization threshold, $E_e$ is initial electron energy. (b)~Energy level of the $^{40}$Ca ion. (c)~LIF spectrum of the plasma ions obtained by scanning the 397~nm laser frequency. The peak signal corresponds to a zero detuning $\Delta_{397}=0$ of the 397~nm from the $^2S_{1/2}$ -- $^2P_{1/2}$ transition frequency. The blue dashed curve corresponds to the best Voigt fit. The filled area under the LIF spectrum curve is proportional to the total number of ions. (d)~Typical fluorescence image of a stationary ion cloud using an exposure time of 1~s and fluorescence spatial distribution profile of ions along the horizontal axis.}
\end{figure}

The experimental procedure is as follows. First, we cool about $5\cdot10^7$ of $^{40}$Ca atoms to $T_a=3$~mK, with peak density of about $n_{MOT}=7\cdot10^9$~cm$^{-3}$ in a conventional MOT~\cite{zelenerCoherentExcitationRydberg2018}. The atomic cloud in our experiment is approximately Gaussian $n(x,y,z)=n_0\cdot e^{(-x^2/2\sigma_x^2)}e^{(-y^2/2\sigma_y^2)}e^{(-z^2/2\sigma_z^2)}$ with size along the horizontal axes equal to $\sigma_x=\sigma_y=0.98(1)$~mm and along vertical axis, to $\sigma_z=0.72(1)$~mm. For the MOT we use a 423~nm cw cooling laser driving the transition $4s^2 \:^1S_0$ -- $4s4p \: ^1P_1$ and,~the~672~nm cw re-pump laser to prevent atomic losses to the $3d4s \: ^1D_2$ state. The continuous coupling of calcium atoms to plasma is performed by switching on the 390~nm ionizing laser beam, which continuously ionizes $4s4p \: ^1P_1$ atoms in the MOT. The 390~nm laser peak intensity $I_{390}$ is in the range from 66 to 153~mW/cm$^2$ (Fig.~\ref{fig:1}(a)). The frequency of the 390~nm laser is tuned just above the ionization threshold $E_I$ for producing electrons whose excess kinetic energy $E_e$ vary in range from 0 to 15~K. 

Since we deal with alkaline-earth calcium atoms in the MOT, we can apply the laser-induced fluorescence (LIF) method~\cite{mcquillenImagingEvolutionUltracold2013} in order to observe in our experiment the Ca ions of the plasma at a wavelength 397~nm in the visual spectrum. We obtain the LIF of $^{40}$Ca ions using a probe cw laser having a wavelength of 397~nm and the peak intensity of 212~mW/cm$^{2}$ and an optical re-pump cw laser having a wavelength of 866~nm and the peak intensity of 6.8~mW/cm$^2$ (Fig.~\ref{fig:1}(b)). In all the experiments, the size of the LIF laser beam is made significantly larger than the size of the MOT cloud, for good spatial overlapping of ions and laser beams. 

The LIF of the SUCP ions and the MOT atoms is focused into an optical fiber using a lens. After passing the optical fiber, the fluorescence of atoms and ions is separated by means of a diffraction grating and is recorded using photomultiplier tubes. An example of the LIF spectrum of experimental ions is shown in the Fig.~\ref{fig:1}(c). Fluorescence image of a stationary ion cloud using an exposure time 1~s is shown in the Fig.~\ref{fig:1}(d).

We also record the spatial distribution of the LIF imaging of the ions by means of a band pass filter and a CCD camera. The ion space distribution is well approximated by a Gaussian distribution having sizes along the horizontal axis equal to 0.83(1)~mm and, along the vertical axis, to 0.75(1)~mm. 

\begin{figure}
  \includegraphics[width=0.5\textwidth]{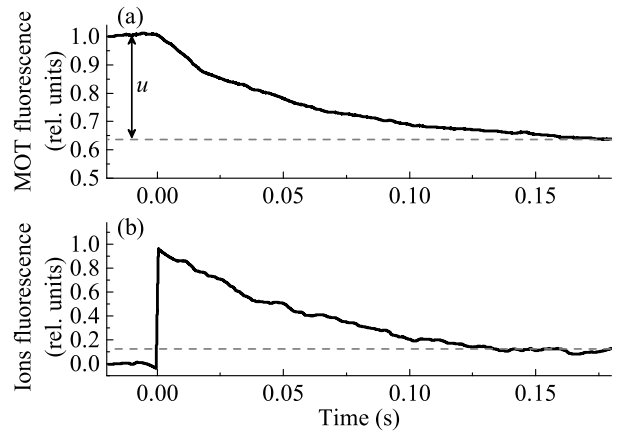}
  \caption{\label{fig:2} (a) The dependence of LIF of MOT signal on time. $u$ is the relative reduction of the MOT atomic fluorescence signal after the ionizing laser is switched on. (b) The dependence of the peak LIF of ions signal on time. A established MOT and ions fluorescence signal level after the ionizing laser beam is turned on is shown with dashed lines.}
\end{figure}

When the ionizing laser beam is turned on, we observe a transient process before a steady-state ions LIF signal is established. Figure~\ref{fig:2} shows the dependence of the peak $^2S_{1/2}$ -- $^2P_{1/2}$ LIF signal (i.e. the signal at a zero detuning of the 397~nm laser frequency ($\Delta_{397}=0$) from the $^2S_{1/2}$ -- $^2P_{1/2}$ transition frequency) on time. Within about 200~µs, the LIF signal reaches its maximum value and then, during about 130~ms, it approaches its steady-state value. We also note that the fluorescence of the MOT atoms decreases after the ionization laser is switched on since the ionization leads to additional atom losses from MOT.

\begin{figure}
  \includegraphics[width=0.5\textwidth]{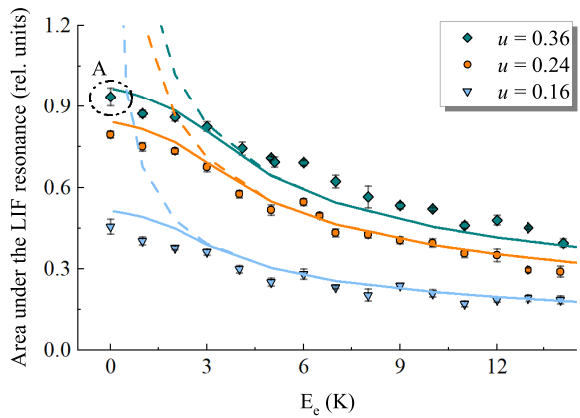}
  \caption{\label{fig:3} The dependence of area under the LIF spectrum curve on the initial electron energy. Points are the experimental data, solid curves are the scaled theoretical calculations taking into account three-body recombination (TBR) and dashed curves are the analytical calculations without the TBR. $u$ is the relative reduction of the MOT fluorescence after the ionizing laser is switched on. $u=0.36, 0.24, 0.16$ corresponds to peak intensity of 390~nm laser 153, 109, 66~mW/cm$^2$ accordingly. ``A'' is the experimental point which corresponds to maximum ions peak density of about $6\cdot10^5$~cm$^{-3}$.}
\end{figure}

In our experiment, we cannot directly measure the density of SUCP ions and their number using the absorption imaging method due to the low signal-to-noise ratio. However, we can make a relative measurement of the plasma ions number via integrating the LIF spectrum, since the area below the LIF spectrum curve is proportional to the total number of ions. Figure~\ref{fig:3} shows the dependence of the area under the LIF spectrum curve on the initial electron energy $E_e$ when the steady-state mode is reached for 3 different ionization laser intensities: 153, 109, 66~mW/cm$^2$. Each experimental point is averaged over about 10--15 different area measurements, each of which was obtained from the ion fluorescence curve averaged over about 30 frequency scans. The results obtained can be qualitatively described by the following simple model.

First, on the basis of the rate equations for atom kinetics in the MOT in a steady-state mode, we obtain the rate of the ions generation:
\begin{eqnarray}
  S=(2\pi)^{3/2}\sigma_x^2\sigma_zun_{MOT}^0/\tau_{MOT},
\end{eqnarray}
where $n_{MOT}^0$ is the MOT atom peak density before the ionization laser is switched on, $\tau_{MOT}=0.18$~s is a lifetime of atoms in the MOT (measured by turning off the Zeeman slower), $u$ is the relative reduction of the MOT atomic fluorescence signal after the ionizing laser is switched on. The value of $u$ is uniquely associated with the ionization laser intensity. 

Further, we note that in the steady-state mode, the above-defined rate of the ions generation $S$ should be equal to total ion losses. The losses include the hydrodynamic ion outflux and the recombination of ions and electrons. The ionic outflux is generated by the charge imbalance which occurs due to fast electrons leaving the plasma. After some electrons leave, the remained electrons are trapped in the excessive positive ion charge which, in turn, form the radial electric field accelerating the ions themselves. Assuming spherical symmetry for qualitative analysis we can estimate the potential due to the charge imbalance as 
\begin{eqnarray}
  \frac{e^2(N_i-N_e)}{4\pi\varepsilon_0R_{eff}}\sim\frac{3}{2}k_BT_e,
\end{eqnarray}
where $N_i$ and $N_e$ are numbers of ions and electrons in the plasma cloud, $k_B$ is the Boltzmann constant, $T_e$ is the electron temperature, $R_{eff}$ is some effective size of the plasma cloud (here and further all the equations are in SI units). From this we can deduce that the ionic kinetic energy should be proportional to the potential depth, therefore, their characteristic hydrodynamic velocity is $v\sim\sqrt{k_BT_e/M_i}$ where $M_i$ is the ion mass. Then for the qualitative analysis we estimate ionic flux losses in ions per second as $N_iv/R_{eff}$. 

In an ultracold plasma, the main recombination process is the three-body recombination (TBR). The rate of ionic losses due to the TBR is $\nu_{TBR}=C_Rn_e^2/(k_BT_e )^{9/2}$, where $n_e$ is the electron density in~cm$^{-3}$, $C_R=3.5\cdot e^{10}/((4\pi\varepsilon_0)^5\sqrt{m_e})$ is the TBR constant~\cite{zeldovichPhysicsShockWaves2002} ($m_e$ is the electron mass). By equating the ion source to the losses $S\sim N_iv/R_{eff}+N_i\nu_{TBR}$, we can estimate the steady-state ion number as: 
\begin{eqnarray}
  N_i=S\left(C\frac{\nu}{R_{eff}}+\nu_{TBR}\right)^{-1},
\end{eqnarray}
where $C$ is the fitting parameter.

As is well known~\cite{robicheauxSimulationExpansionUltracold2002}, the electron temperature in an ultracold plasma is strongly affected by the heating due to the TBR process. In our simple analysis we assume that the steady-state value of $T_e$ is established due to an interplay of the TBR heating and the constant influx of cold electrons formed in the process of continuous ionization of the cold atomic cloud (we assume that the electron source per second is the same as that for ions, i.e. $S$):
\begin{eqnarray}
  k_B T_e=\frac{2}{3}\nu_{TBR}\frac{N_e}{S}E^*+k_B T_K,
\end{eqnarray}
where $T_K=\frac{2E_e}{3k_B}$ is the initial kinetic energy of electrons defined by the ionization laser detuning above the ionization threshold and, $E^*$ is the energy which is transferred to free electrons per one recombination act.

In order to estimate the energy yield per recombination act, we use the expression~\cite{zeldovichPhysicsShockWaves2002,Kusnetsov1965}:
\begin{eqnarray}
  E^*=3.1\cdot10^{-5}E_In_e^{1/6}T_e^{1/12},
\end{eqnarray}
where $E_I$ is $^{40}$Ca atom ionization potential. 

Combining the solutions of the above equations and substituting $n_e=N_e/(\frac{4}{3}\pi R_{eff}^3)$ for electron number density we find the steady-state values of $N_i$ and $T_e$ for various initial electron energies $T_K$ and for various source term values $S$. Since the total ion number is proportional to the area under the LIF spectrum curve of ions, we plot the dependence of calculated values of $N_i$ on $E_e$ in Fig.~\ref{fig:3}, scaled to the area values (divided by 17000). In order to calculate $N_i$ we use $C=1$ and the effective plasma cloud size $R_{eff}=0.18$~cm (which is close to the actual ionic cloud size). Although the above theory pretends only to qualitatively describe the experiment, the agreement of the theoretical model and the experimental results is quite good. In Fig.~\ref{fig:3} also shown results of calculations without the account of the TBR process. In our experiments the TBR is significant for $T_e<2$~K.

The above-said theory can be applied for making an indirect estimation of the ion number, density and electron temperature of our plasma. As a result, we estimate that the total number of ions reaches $N_i\approx 16000$ for the experiment A in Fig.~\ref{fig:3}. The peak density of ions for this experiment can be estimated as $n_i\approx 6.6\cdot10^5$~cm$^{-3}$ at the minimum steady-state electron temperature of $T_e\approx 2$~K which corresponds to the initial electron energy $T_K=0$~K. At these parameters, the Debye length $\lambda_D=\sqrt{\varepsilon k_B T_e/n_e e^2}\approx 0.012$~cm is well below the system size, so the plasma approximation can be used with a good accuracy even for the lowest electron temperature considered.

Now we can estimate the coupling strength of our plasma, which is characterized by a coupling parameter. The coupling parameter is usually defined as $\Gamma_\alpha=e^2/(4\pi a \varepsilon_0k_BT_\alpha)$. Here $\alpha$ corresponds to the particle species ($e$ is for electrons or $i$ is for ions), $T_\alpha$ and $n_\alpha$ are the temperature and the number density of charged particles of the type $\alpha$, $a=(3/4\pi n_\alpha)^{1/3}$ is the radius of the Wigner-Seitz cell. For the particles density $n_i\approx 6\cdot 10^5$~cm$^{-3}$ and the temperature $T_e=2$~K, the estimated coupling parameter $\Gamma_e$ is about 0.1, which corresponds to moderate coupling. In a plasma, the parameter $\Gamma_e=0.1$ corresponds to that beyond which the plasma cannot be treated as ideal gas and, Coulomb interaction of the particles should be taken into account. 

The ions are accelerated in the excessive positive charge and their kinetic energies at the plasma edge are governed by the electron temperature. However, in the bulk of the plasma, the ion temperature is low because of ongoing ionization of the cold atoms. Following~\cite{bergesonExploringCrossoverHighenergydensity2019}, we estimate that the temperature of ions is within the limits determined by the disorder-induced heating (DIH) effect $T_i=T_{DIH}\approx e^2/(2.3\cdot 4\pi a \varepsilon_0 k_B)$. This temperature corresponds to the ion coupling parameter $\Gamma_i\approx 2.3$ in the bulk of the plasma cloud.

\begin{figure}
  \includegraphics[width=0.5\textwidth]{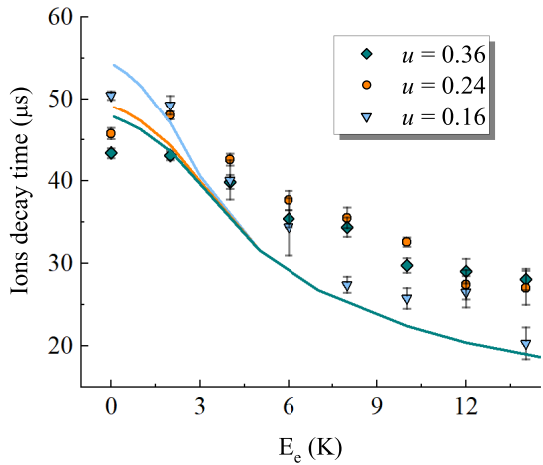}
  \caption{\label{fig:4} Decay time of the plasma cloud versus the initial energy of free electrons. Points are the experimental data; solid curves are the analytical calculations taking into account TBR. Values of relative reduction of the MOT fluorescence $u$ are the same as in Fig.~\ref{fig:3}}
\end{figure}

We also study dynamical properties of the plasma and measure the plasma decay time, i.e. the time during which the ions LIF signal is reduced to the level of $1/e$ after the ionizing laser beam is switched off. In Fig.~\ref{fig:4} the decay time is plotted for various initial electron energies and various ionization laser intensities. Each point on the graph is an average over about 3000~cycles of switching off the ionization laser. It can be seen that with a decrease in the electron temperature, the decay time of the ions cloud increases rapidly, but is limited when $E_e$ is approaching zero due to the TBR heating.

For comparing our theoretical calculations with the experimental results, we estimate the theoretical decay time of ions cloud as 
\begin{eqnarray}
  \tau=\left(\frac{\nu}{R_{eff}}+\nu_{TBR}\right)^{-1}.
\end{eqnarray}

As can be seen from Fig.~\ref{fig:4}, the agreement of the theory and experiment for the decay time is good as well. 

In conclusion, we were able to experimentally produce a strongly coupled steady-state ultracold plasma by ionization of a cold atomic cloud in the operating MOT by a cw laser radiation. The dependencies of the plasma properties on initial electron energies and ionization laser intensities are in good agreement with simple theoretical model. The proposed theory is qualitative only, it does not take into account spatial inhomogeneity, it also does not take into account the presence of the magnetic field of the working MOT, which is about 30~G/cm. As shown in recent studies, the presence of a magnetic field in an ultracold plasma can lead to a significant decrease of the TBR rate and the DIH rate~\cite{bobrovElectronDistributionFunction2013,tiwariReductionElectronHeating2018}. It is also shown that a magnetic field with an anti-Helmholtz geometry is a trap for the ultracold plasma~\cite{gormanMagneticConfinementUltracold2021}. For a more rigorous description of the plasma, we plan to further study the steady-state plasma formation by means of the molecular dynamics method with full account for the magnetic field. 

We also plan to increase the magnetic field in our MOT in order to achieve magnetic confinement in the same fashion as it was made in~\cite{gormanMagneticConfinementUltracold2021}, but in steady-state conditions. We think that the steady-state ultracold plasma can be an excellent model object for studying properties of long–lived strongly coupled systems.

\begin{acknowledgments}
  The research is supported by the Russian Science Foundation, Grant No.~18-12-00424.
\end{acknowledgments}


\bibliography{BiblioZotero2}

\end{document}